\documentclass[twocolumn,twoside,slac_two]{revtex4}
\usepackage{graphicx}
\usepackage{fancyhdr}
\begin{document}

\title{The HAWC observatory as a GRB detector}


\author{D. Zaborov$^{1}$ for the HAWC collaboration}
\affiliation{$^{1}$Department of Physics, Pennsylvania State University, 104 Davey Laboratory, University Park, PA 16802, USA}

\begin{abstract}
The High Altitude Water Cherenkov Observatory (HAWC) is an air shower array currently under construction in Mexico at an altitude of 4100 m.
HAWC will consist of 300 large water tanks covering an area of about 22000 square meters and instrumented with 4 photomultipliers each.
The experimental design allows for highly efficient detection of photon-induced air showers in the TeV and sub-TeV range and gamma-hadron separation.
We show that HAWC has a reasonable chance to observe the high-energy power law components of GRBs that extend to $\approx$50 GeV.
In particular, HAWC will be capable of observing events similar to GRB~090510 and GRB~090902B.
The observations (or non-observations) of GRBs by HAWC will provide information on the high-energy spectra of GRBs.
An engineering array consisting of 6 water tanks was operated at the HAWC site since September 2011, collecting 3 months of data.
An upper limit on high energy emission from GRB~111016B is derived from these data.

\end{abstract}

\maketitle

\thispagestyle{fancy}


\section{Introduction}
Recent observations by Fermi LAT suggest that the high-energy emission of some GRBs extends at least to 30 GeV. 
In particular, a hard power law component was detected in the spectra of 
the long GRB~090902B \cite{grb090902b} and the short GRB~090510 \cite{grb090510}.
The highest energy photon recorded from GRB~090902B was 33 GeV, or 94 GeV corrected for redshift.
The high energy component, 
not described by the Band function \cite{Band1993}, 
is currently a challenge to GRB models.
During the propagation of gamma rays through the interstellar medium,
an attenuation due to interactions with the
extra-galactic background light (EBL) is expected.
Consequently, a spectral energy cutoff can be a probe for the EBL density
or source properties (e.g. bulk Lorentz factor).
The extension of the observations to higher energies requires a detector with a large effective area.
This can be provided by Imaging Atmospheric Cherenkov Telescopes (IACTs)
and Extensive Air Shower (EAS) arrays.
The chances of prompt observations of a GRB by IACTs are, however, limited by the small duty cycle ($\approx$ 10\%)
and restricted field of view (5$^\circ$ in diameter or less).
The EAS arrays provide a large field of view ($\approx 2$~sr) and near 100\% duty cycle.
An energy threshold $<$~100 GeV can be achieved by placing the array at a high altitude (closer to the air shower maximum).
So far the searches for GRBs with EAS arrays have not yielded positive detections
due to insufficient sensitivity of the existing arrays 
(although tentative detections have been claimed by several experiments, e.g. Milagrito \cite{atkins03}).

\section{The HAWC observatory}
\begin{figure}
  \centering
  \includegraphics[width=0.98\linewidth]{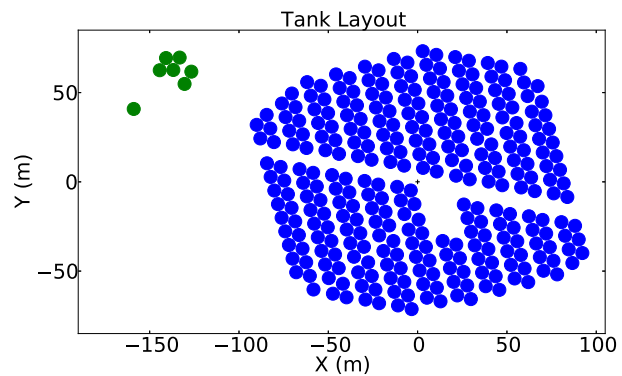}
  \caption{HAWC layout. The group of six WCDs
           at the top left correspond to VAMOS. The electronics
           counting house is at the empty region in the center
           of the array.}
  \label{fig:layout}
\end{figure}

The High Altitude Water Cherenkov (HAWC) observatory is a very
high-energy ($\sim$TeV)~gamma-ray detector currently under construction near
the peak of Volc\'an Sierra Negra, Mexico. HAWC is located at 4100 m altitude,
N $18^\circ 59^\prime 48^{\prime \prime}$, W $97^\circ 18^\prime 34^{\prime \prime}$.
When completed in 2014, HAWC will consist of 300
water Cherenkov detectors (WCD) of 7.3 m diameter and 4.5 m depth,
covering an area of about 22000~m$^2$ (see Fig.~\ref{fig:layout}).
Four large photomultiplier tubes (PMTs) will be located near the bottom of each WCD.\footnote{The effective
area and sensitivity estimates presented below correspond to simulations of only three PMTs per WCD and are therefore conservative.}
The PMTs capture Cherenkov light produced in water by the charged particles 
that compose an extensive air shower.
The main data acquisition system (DAQ) consists of 
front-end electronics boards, % that receive the PMT signals %, which are installed in the counting house,
time-to-digital converters (TDC) and computer systems that perform the readout
and apply trigger (in triggered mode).
Hit arrival times are used to reconstruct the incident direction of the shower.
The rejection of hadronic showers relies on the shower lateral size and high amplitude pulses produced by muons.

\begin{figure}
  \centering
  \includegraphics[width=0.99\linewidth,height=65.0mm]{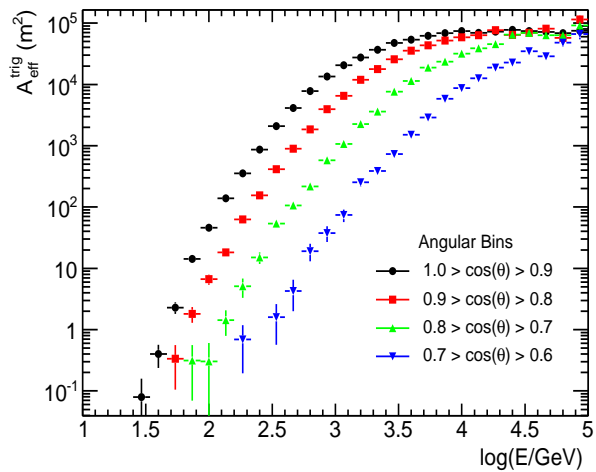}
  \caption{Effective area of HAWC for photon-induced showers.
           A~trigger threshold of 70 PMT hits is assumed.
           Showers reconstructed with $> 0.8^\circ$ error are excluded.
           No gamma-hadron separation cut is applied.
          }
  \label{fig:effarea}
\end{figure}

The effective area of HAWC will approach $\approx$10$^5$~m$^2$ at $E > 3$ TeV (see Fig.~\ref{fig:effarea}).
At $E < 1$ TeV the effective area falls rapidly with decreasing energy
%due to the absorption of the air shower particles in the atmosphere.
due to the majority of the air showers not reaching the ground.
Below 30 GeV HAWC is generally less sensitive than satellites such as Fermi LAT.
An angular resolution of 0.1$^\circ$ can be achieved at $E > 5$ TeV.

Scalers, the second data acquisition system, will measure PMT counting rates.
A short transient flux of gamma rays can be identified as a statistical excess in PMT rate across the detector.
This method provides an energy threshold of a few~GeV.

A test array of six WCDs, called VAMOS (Verification And Measuring of Observatory
System), was built at the HAWC site in 2011. The WCDs were
filled with water and instrumented with 4 to 7 PMTs per WCD. 
Over 3 months of raw untriggered data have been collected with this array.
Construction of the main HAWC array started in February 2012,
with first 30 WCDs delivered in September.

\section{Sensitivity of HAWC to Gamma Ray Bursts}
\begin{figure}
  \centering
  \includegraphics[width=0.99\linewidth]{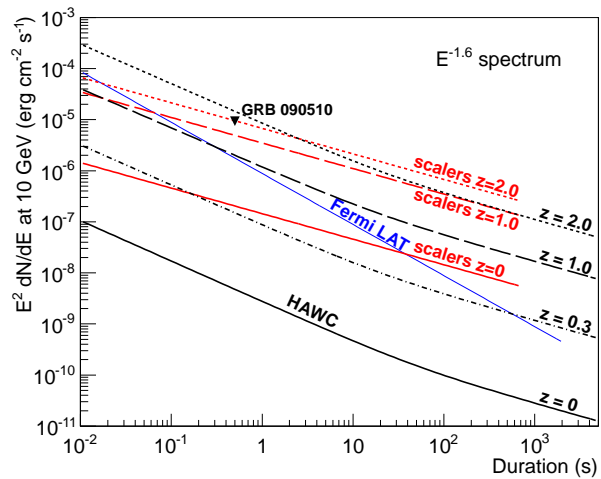}
  \caption{Sensitivity ($5\, \sigma$ discovery potential) of HAWC 
           using the main DAQ and the scaler DAQ as a function of burst duration.
           The source zenith angle is set to 20$^\circ$.
           The Gilmore model of gamma ray attenuation by EBL \cite{gilmore09}
           is used to obtain the sensitivity curves for different redshifts.
           Also shown is the flux necessary for the observation of 1 photon above 10 GeV by Fermi LAT.
          }
  \label{fig:sensitivity}
\end{figure}

The sensitivity of HAWC to GRBs is illustrated in Fig.~\ref{fig:sensitivity}.
It is assumed that the GRB search time window and position on the sky 
can be set based on observations in other wavelengths (e.g. by satellites).
As can be seen, the sensitivity of HAWC is comparable to Fermi LAT's sensitivity above 10 GeV,
but depends strongly on the redshift (spectral cutoff).
HAWC scalers complement the main DAQ, covering short GRBs with soft spectra and high redshifts (cutoffs at $E<100$ GeV). 
The brightest GRBs detected by Fermi, such as GRB~090510, should be observable with HAWC if the cutoff is above 50-60 GeV.
The sensitivity estimates presented here are based on conservative assumptions
for the trigger threshold (for the main DAQ) and the number of active PMTs.
For a detailed report on HAWC sensitivity to GRBs see \cite{hawcGRBsensi}.

\section{Limit on high energy emission from GRB 111016b}
The data collected with VAMOS can be used to search for high energy (HE) emission from GRBs.
Due to its small size (31 PMTs versus 1200 planned for HAWC),
VAMOS only provides $\approx 15$\% of the sensitivity of the full array
when using the scalers DAQ (sensitivity runs as $1/\sqrt{N_{PMT}}$)
and $\approx 3$\% for the main DAQ analysis
(where angular resolution and gamma-hadron separation have a significant effect).
Nevertheless it is comparable in sensitivity to Milagro, 
a previous generation experiment that operated at a lower altitude of 2630 m
(see e.g. \cite{milagro060427b}).
%(see e.g. \cite{milagro2006}).

\begin{figure}
  \centering
  \includegraphics[width=0.98\linewidth]{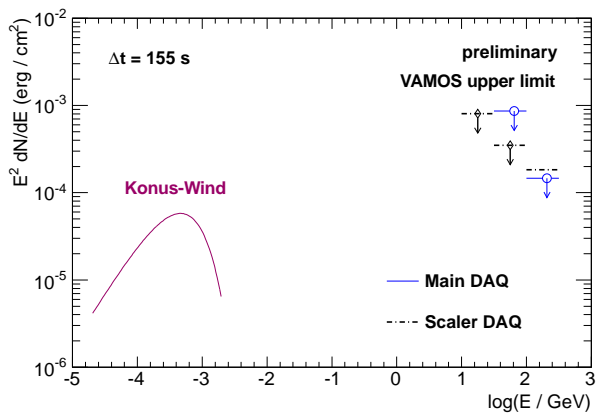}
  \caption{Upper limits on high energy emission from GRB 111016B imposed by VAMOS data.
           Two sets of limits are shown: using reconstructed air shower events (main DAQ) and using PMT rates (scaler DAQ).
           The main DAQ (scaler DAQ) limits are given for two (three) energy bands separated by half a decade in energy.
           All limits are at 90\% confidence level.
           The spectral fit reported by Konus-Wind (GCN circular 12456) is shown for comparison.}
  \label{fig:vamos_limit_grb111016b}
\end{figure}

A search for HE emission has been performed using VAMOS data for a long-duration, intense GRB~111016B
detected by the IPN network (GCN circular 12452). % \cite{GCN12452}.
The GRB occurred at a zenith angle of 32$^{\circ}$ in the VAMOS field of view.
No redshift information is available for this GRB.
The number of air showers detected during a 155~s time interval around the GRB
(including 5 s before T0 = 22:41:40 UT plus the reported GRB duration)
and reconstructed within 6$^{\circ}$ from the GRB position
was compared to the background estimate 
based on the event rate in the same angular bin during a 7 hr period including the GRB.
A negative fluctuation of $\approx\,0.6\,\sigma$ was found.
A 90\% C.L. upper limit on the number of signal events was derived following the method of Feldman and Cousins~\cite{FeldmanCousins1998}.
The limit was then converted to flux units using a Monte Carlo simulation of the detector response.
The limits computed for two different energy bands are presented in Fig.~\ref{fig:vamos_limit_grb111016b}.
Assuming a power law spectrum with a cutoff at 100 GeV the upper limit on E$^2$ dN/dE at 65~GeV is 8.6 $\cdot$ 10$^{-4}$ erg/cm$^2$.
For a spectrum extending up to 316~GeV, the limit on the $>$100 GeV emission is
1.5 $\cdot$ 10$^{-4}$ erg/cm$^2$ at 208 GeV.

A second set of upper limits
was obtained for GRB~111016B using the PMT rate data recorded by VAMOS scalers.
The number of PMT counts in the 155~s time window was compared to
the 7 hr of data surrounding the GRB.
It was found that the rate had an underfluctation of $1 \sigma$ in the 155 s window.
A 90\% C.L. upper limit was derived using the same method as for the main DAQ.
The flux limit computed for three energy bands is shown in Fig.~\ref{fig:vamos_limit_grb111016b}.
As one can see, at E $>$ 100 GeV more stringent limits come from the air shower analysis,
while the scalers technique provides additional sensitivity to lower energies.

Two additional time windows were defined to search for HE emission 
during the main prompt emission peak recorded by Konus-Wind (30 s starting at T0)
and extended emission (315 s starting T0 - 15 s).
The measurements by both DAQ systems were found to be within $\approx1.5\sigma$
from the background estimates in all cases.

\bigskip % extra skip inserted
\section{Conclusion}

HAWC is a new generation wide field of view gamma-ray telescope currently under construction in Mexico.
The high altitude, high duty cycle and large field of view make HAWC a suitable detector for gamma-ray bursts.
HAWC will provide a realistic opportunity to observe
the high-energy power law components of GRBs that extend at least up to 50-60 GeV.
HAWC measurements will provide valuable information on the high-energy cutoff 
in the intrinsic GRB spectra and/or EBL absorption cutoff.
An engineering array consisting of six WCDs has been operated at the HAWC site since September 2011.
The collected data were used to set an upper limit on high energy emission from GRB~111016B.

\bigskip % extra skip inserted
\begin{acknowledgments}
The HAWC project is supported by the National Science Foundation, the US Department of Energy Office of High-Energy Physics,
the LDRD program of Los Alamos National Laboratory and the University of Wisconsin Alumni Research Foundation, in U.S.;
Consejo Nacional de Ciencia y Tecnolog\'{\i}a, Red de F\'{\i}sica de Altas Energ\'{\i}as, DGAPA-UNAM and VIEP-BUAP, in Mexico.
\end{acknowledgments}

\bigskip % extra skip inserted

\end{document}